\documentclass[11pt]{article} 
\usepackage{epsfig,a4}
\usepackage[intlimits]{amsmath}
\usepackage{amssymb}
\usepackage{psfrag}
\usepackage{graphicx}
\def\be{\begin{equation}}
\def\ee{\end{equation}}
\def\bea{\begin{eqnarray}}
\def\eea{\end{eqnarray}}

\def\fr{\frac}

\def\del{\partial}
\newcommand\lbr{\left(}
\newcommand\rbr{\right)}

\def\l.{\left.}
\def\r.{\right.}
\newcommand\qslash{\!\not \! q}

\begin{document}
\begin{titlepage}
\begin{flushright}
IFUM-857-FT\\
ECT$^*$-05-20\\
hep-ph/0601271
\\
\end{flushright}
\vfill
\begin{center}
\boldmath
{\LARGE{\bf Gribov's Picture of Confinement and}}
\\[.3cm]
{\LARGE{\bf Chiral Symmetry Breaking}}
\unboldmath
\end{center}
\vspace{1.2cm}
\begin{center}
{\bf \Large
Carlo Ewerz
}
\end{center}
\vspace{.2cm}
\begin{center}
{\sl
Dipartimento di Fisica, Universit{\`a} di Milano and INFN, Sezione di Milano\\
Via Celoria 16, I-20133 Milano, Italy\\
\hspace{1cm}\\
ECT$^*$, Strada delle Tabarelle 286, I-38050 Villazzano (Trento), Italy\\
\hspace{1cm}\\
email: Ewerz@ect.it
}
\end{center}
\vfill
\begin{abstract}
\noindent
Gribov's scenario of confinement caused by supercritical charges is 
described and the present status of Gribov's equation for the 
Green function of light quarks is discussed. 
\end{abstract}
\vspace*{3cm}
\vfill
\end{titlepage}

\section{Introduction}
\label{sec:intro}

One of the most challenging problems in quantum field theory is 
to understand the confinement mechanism in QCD. 
V.\,N.\ Gribov attacked this problem in a longlasting effort in  
which he developed both a physical picture and a quantitative approach 
to describing the confinement mechanism. 
His main concern was not just to find {\sl a} confinement mechanism but 
rather {\sl the} confinement mechanism relevant to the real world in 
which we live. It was a key observation for him \cite{Gribov:1986vp} 
that the existence of 
very light quarks can drastically change the vacuum structure of QCD. 
Therefore a solution of the confinement problem appeared impossible to 
him without understanding the effects of light quarks. From this point 
of view the question of confinement in pure gluodynamics is interesting 
but probably irrelevant for our real world. 

From the early 90's onwards the problem of light quarks therefore became 
the main focus of Gribov's research \cite{Gribov:1991rv}. 
In this context he addressed several 
key questions. One of them is the way in which the Pauli principle can 
provide a binding mechanism even for almost massless quarks due to 
the occurrence of supercritical charges. Another 
one is the relation between the infrared and ultraviolet regimes of the 
theory which appear to be linked much more closely than one might 
naively expect. He also addressed the question whether and how 
the perturbative expansion has to be reconsidered in a theory in which the 
fundamental fields of the Lagrangian do not exist as asymptotic states. 
This is in fact a very deep problem as one recognizes by recalling that 
the usual treatment of propagators with the Feynman 
$i \epsilon$-prescription requires that the perturbative vacuum is stable -- 
which is not 
true in QCD where quarks and gluons cannot propagate freely. 

In 1997 V.\,N.\ Gribov was convinced that he had finally found answers 
to these questions and that he understood the physics 
of light quarks sufficiently well. He then collected his results in two 
papers \cite{Gribov:1998kb,Gribov:1999ui} which were supposed to conclude 
his study of confinement in QCD. Unfortunately, the second paper had to 
remain unfinished when he passed away. 

In the present contribution we describe the idea of supercritical charges 
in QCD, give a review of the Gribov equation for the propagator of 
light quarks, and summarize the present status of that equation. 
For a more detailed exposition of the motivations and physical ideas 
behind Gribov's approach we refer the reader to the recent 
review \cite{Dokshitzer:2004ie}.  
All of Gribov's original papers on confinement (as well as many of his 
papers and lectures on high energy scattering) are collected in the 
book \cite{Gribov:phasis}. 

\section{Light Quarks and Supercritical Charges}
\label{sec:supercritical}

The phenomenon of supercritical charges is well known in QED. 
Consider the energy levels that an electron would have in the 
static external field of a heavy nucleus $A_Z$ with charge $Z$. When 
$Z$ is increased the energy levels are lowered. At $Z>137$ 
the lowest (1s) level sinks below $-m_e c^2$ and hence dives into 
the Dirac sea.\footnote{The critical value $Z_c=137$ holds for 
a pointlike nucleus, in the more realistic situation of an extended 
nucleus it would be somewhat higher.} 
In this situation an electron from the Dirac sea fills this level 
and forms a strongly bound state with the nucleus while a positron 
is emitted. This is also reflected in the quantum mechanical energy 
of the electron as it results from the Dirac equation which becomes 
complex at $Z>137$, signaling an instability that corresponds to 
the so-called `falling onto the center'. In summary, the heavy nucleus 
decays into a supercritically bound state $A_{Z-1}$ with a smaller 
charge $Z-1$ and a positron, 
\be
\label{nucleusdecay}
A_Z \to A_{Z-1} + e^+ \,.
\ee
Starting from a highly charged nucleus that process continues for 
higher energy levels until the resulting bound state is subcritically 
charged. 

Gribov's idea is that a similar mechanism might apply to the color 
charges of QCD if light quarks exist with a Compton wave length 
that is larger than the typical color charge radius or confinement radius. 
He proposed the intriguing possibility that in QCD 
already the color charge of a {\sl single} quark could be supercritical, in 
contrast to QED where a very large charge is required. Note that in QED 
the relevant parameter for the phenomenon of supercritical charges 
is in fact $\alpha_{\mathrm{em}} Z$ and that also a large electromagnetic 
coupling $\alpha_{\mathrm{em}}$ could trigger the instability. 
In QCD the role of the supercritical charge is in fact played by the strong 
coupling constant $\alpha_s$ which becomes large at low momentum 
scales. As we will see in section \ref{sec:csb} below supercritical behavior is 
expected if the strong coupling constant exceeds a critical value 
$\alpha_c$ given by \cite{Gribov:1991rv} 
\be
\label{critcoupling}
\frac{\alpha_c}{\pi} = C_F^{-1} \left( 1 - \sqrt{\frac{2}{3}} \right) 
= 0.137 \,,
\ee
with the Casimir operator $C_F=(N_c^2-1)/(2N_c)=4/3$. 
Note that this critical value is remarkably small. 

Let us now briefly discuss how supercritical color charges change 
the vacuum structure of light quarks. 
(A more detailed account of the confinement mechanism caused 
by light quarks can be found in Refs.\ \cite{Gribov:Orsay,Ewerz:2000qb}.) 
The emerging picture is again most conveniently discussed in the 
language of the Dirac sea. For subcritical coupling we have the 
usual situation with empty positive energy quark states and a filled 
Dirac sea of negative energy states. If the coupling is larger than 
the critical value $\alpha_c$ additional states appear. Supercritical bound 
states can form consisting of a quark and an antiquark which both have 
positive kinetic energy. But due to the supercritical binding, the bound state 
has a {\sl negative} total energy. In order to avoid having infinitely many 
mesonic negative energy states the corresponding quark and antiquark states 
have to be filled in the vacuum. In particular, the vacuum then contains 
occupied quark states of {\sl positive kinetic energy}, in addition to the negative 
energy states in the Dirac sea. In such a vacuum state the Pauli exclusion 
principle prevents single quarks from propagating freely since the 
states required for that are already occupied. In that sense the Pauli 
principle can bind even massless quarks which usually tend to move at 
the speed of light. A single quark put into the QCD vacuum would 
undergo a decay into a supercritically bound meson and another quark, 
$q \to M + q$. Since the resulting quark is again supercritically charged 
that process would continue and eventually shield the color charge completely, 
such that a single quark would only exist as a resonance but not as an 
asymptotic state. 

Obviously, this confinement mechanism based on the Pauli principle 
works only for quarks, that is for 
fermions in general, but does not apply to gluons. It remains to be 
investigated how gluons are affected by the light quarks in such a picture. 
It is conceivable that the confinement of gluons is, at least to some extent, 
caused by their coupling to confining light quarks, and would then be 
an indirect or second-order effect. 

From a technical point of view the situation in QCD is considerably 
more complicated than the 
example of a heavy nucleus in QED. While a heavy quark can be treated 
in a similar way as the heavy nucleus, the supercritical charge of a light 
quark can clearly not be considered as an external field. In view of this 
problem Gribov invented a new approach to the Dyson-Schwinger 
equation for light quarks that we discuss in the next section. 

Interestingly, the dispersive approach \cite{Dokshitzer:1995qm} 
to power corrections to event shape variables in $e^+e^-$ collisions 
and in deep inelastic scattering suggests that the integrated 
strong coupling constant in the infrared region (assuming of course 
that such a concept exists) satisfies 
\be
\label{intofcoupling}
\alpha_0= 
\frac{1}{2\,\mbox{GeV}} 
\int_0^{2\,\mbox{\scriptsize GeV}} \!\alpha_s(k) \, dk
\, \simeq \, 0.5 
\,,
\ee
see Ref.\ \cite{Dokshitzer:2004tp} for a recent review. From 
(\ref{intofcoupling}) one can deduce that the strong coupling becomes 
in fact larger than the critical value (\ref{critcoupling}) at least 
in some range of momenta, but remains comparatively small in most 
if not all of the small-momentum region. This observation raises some 
hopes that -- unexpectedly -- some aspects of low-momentum 
phenomena like for example hadronization might be accessible 
at least qualitatively in a semi-perturbative way. 

\section{Gribov's Equation for the Green Function of Light \\
Quarks in Feynman Gauge}
\label{sec:eq}

Let us now turn to the Gribov equation (or Gribov-Dyson-Schwinger 
equation) for the Green function of light quarks in Feynman 
gauge \cite{Gribov:1998kb}. 
For Gribov this equation was mainly a technical step on the way to a 
quantitative support of the physical picture discussed in the previous 
section. But it can also be viewed as a new approach to the quark's 
Dyson-Schwinger equation which is very interesting already by 
itself, independently of the supercritical confinement scenario. 

The Green functions of QCD satisfy a tower of coupled Dyson-Schwinger 
equations. Since it is virtually impossible to solve this system of integral 
equations as a whole one usually has to resort to simplifications. 
Most current approaches to the Dyson-Schwinger equations are based 
on truncation schemes (based for example on certain topological classes 
of diagrams), but their accuracy is often very difficult to control. 
Gribov discovered that it is possible to attack the Dyson-Schwinger 
equation for the quark's Green function in Feynman gauge 
based on an approximation 
rather than a mere truncation scheme. The guiding principle in this 
approximation scheme is to collect the most singular contributions 
from the infrared momentum region, that is those contributions that 
are expected to cause chiral symmetry breaking and confinement. 

In Feynman gauge the gluon propagator has the general form 
\be
\label{effphot}
D_{\mu\nu}(k) = - \fr{g_{\mu\nu}}{k^2} \,\alpha_s(k^2)
\,,
\ee
which can be interpreted as a nonperturbative definition of the 
strong coupling constant $\alpha_s$ at low momentum scales. 
We now make the (rather mild) assumption that $\alpha_s(k^2)$ 
is not divergent as $k^2\to 0$ and does not vary too strongly. 
For spacelike momenta a typical example for such a coupling is the model 
\be
\label{Kopplungma}
\alpha_s(k^2)  = \fr{4 \pi}{\lbr 11-\fr{2}{3} n_f \rbr \, \ln  
  ( -k^2 / \Lambda^2_{\mbox{\tiny QCD}} +a)} 
\,, 
\ee
which for $a=6$ satisfies (\ref{intofcoupling}) and agrees with the 
perturbative one-loop running at large momenta. It has been 
shown \cite{Ewerz:2000qb} that the qualitative results discussed below 
are independent of the particular model used for $\alpha_s$ at low 
momentum scales. 

One now starts with the Dyson-Schwinger equation for the quark's inverse 
Green function $G^{-1}(q)$ and applies the d'Alembert operator 
$\del^2= \del^\mu \del_\mu$ with 
\be
\label{defdelq}
\del_\mu = \frac{\del}{\del q^\mu}
\,.
\ee
Let us illustrate this procedure for the one-loop self-energy diagram 
\be
\label{diagganzeinfach}
\centerline{\epsfxsize=3cm \epsfbox{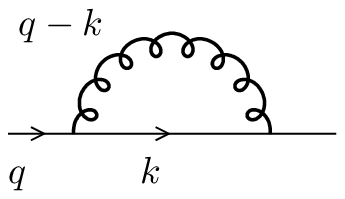}}   
\nonumber
\ee
which is given by 
\be
\label{oneloopselfenergy}
\Sigma = -i \, C_F \frac{\alpha_s}{\pi}
\int \frac{d^4 k}{4\pi^2} 
\, \gamma^\mu G(k) \gamma_\mu \, \frac{1}{(q-k)^2}
\,.
\ee
Firstly, differentiating twice with respect to the external momentum 
$q_\mu$ makes this integral convergent. Secondly, since $1/q^2$ is 
the Green function of the four-dimensional d'Alembert operator 
$\del^2$, 
\be
\label{deltafunktion}
\del^2 \fr{1}{(q-q')^2 +i \epsilon} = - 4 \pi^2 i \,\delta^{(4)}(q-q') 
\,,
\ee
we can eliminate the integral in (\ref{oneloopselfenergy}) and 
obtain to first order in $\alpha_s$ 
\be
\label{gllowestorder}
\del^2 G^{-1}(q) = C_F \frac{\alpha_s}{\pi} \gamma^\mu G(k) \gamma_\mu 
\,.
\ee
One can now systematically proceed to higher order diagrams. In each 
diagram the most singular contribution from the infrared is obtained 
when both derivatives are applied to the same gluon 
line, since only those terms give rise to a delta function according to 
(\ref{deltafunktion}). All other terms are clearly less singular and 
still involve integrals over loop momenta. 

Including higher orders one builds up two full quark-gluon vertices 
$\Gamma_\mu$ which due to the delta functions have to be taken at 
vanishing gluon momentum. Note that as a consequence of the differentiations 
we obtain two full vertices in spite of the fact that the original 
Dyson-Schwinger equation involves one bare and one full vertex. 
The two zero-momentum vertex functions $\Gamma_\mu(q,q,0)$ 
can be related to the inverse Green function via the Ward identity 
\be
\label{wardid}
\Gamma_\mu(q,q,0) = \del_\mu G^{-1}(q) \,,
\ee
and the corrections to the Ward identity due to the nonabelian structure 
of QCD can be argued to be subleading in the Dyson-Schwinger equation 
in the approximation scheme used here. We hence obtain the Gribov 
equation for the light quark's Green function in Feynman gauge, 
\be
\label{gribgl}
\del^2 G^{-1} = g \,(\del^\mu G^{-1}) \,G \,(\del_\mu G^{-1}) 
+ \cdots 
\,,
\ee
where 
\be
\label{QCDdefg}
g = C_F \fr{\alpha_s(q)}{\pi} \,. 
\ee
and the ellipsis in (\ref{gribgl}) stands for less IR-singular terms 
of order $g^2$ which can -- at least in principle -- be systematically 
computed. 
Note that via (\ref{QCDdefg}) we have inserted the running strong 
coupling constant $\alpha_s(q)$ in the Gribov equation. Following 
the derivation outlined above one would first obtain the coupling at 
zero momentum due to the delta function (\ref{deltafunktion}), 
but the difference between the two is subleading in the sense of 
our approximation. 

With the running coupling the Gribov equation has the remarkable 
property that it can reproduce the usual renormalization group 
behavior for the quark propagator in the perturbative region of 
large spacelike momenta. Therefore the equation collects the most 
important terms both in the infrared as well as in the ultraviolet, 
and hence has a good chance of giving a valid description of the 
quark propagator for all momenta. 

The derivation of the equation strongly relies on the mathematical 
identity (\ref{deltafunktion}) and hence on the Feynman gauge. 
Most other approaches to the Dyson-Schwinger equations use 
Landau gauge such that a direct comparison with those approaches 
is rather difficult. If one succeeds in deriving a genuinely gauge-independent 
phenomenon like chiral symmetry breaking or confinement in one 
gauge, however, it will clearly hold in any gauge. 

We finally point out that the same equation (\ref{gribgl}) applies 
also for electrons in QED after replacing $C_F$ by $1$ and $\alpha_s$ 
by $\alpha_{\mathrm{em}}$. However, the behavior of the coupling 
constants and the boundary conditions are very different in the 
two theories. 

\section{Chiral Symmetry Breaking}
\label{sec:csb}

On general grounds the quark's Green function has the form 
\be
\label{genparazm}
G^{-1}(q^2) = Z^{-1} ( \, \qslash - M )
\,,
\ee
where $M(q^2)$ is the dynamical mass function of the quark and 
$Z(q^2)$ stands for the wave function renormalization. 

In QCD the infrared and ultraviolet limits of Green functions 
are very closely related \cite{Gribov:1998kb}. 
In order to see this let us consider the limit of vanishing 
coupling $\alpha_s \to 0$, 
corresponding to large spacelike momenta, $-q^2 \to  \infty$, for 
which we obtain the free equation $\del^2 G^{-1}=0$. 
One readily verifies that this equation has the solution 
\be
\label{asymptfourterm}
G^{-1}(q) = Z^{-1} 
\left( \qslash - m - \frac{\nu_1^3}{q^2} 
- \frac{\nu_2^4}{q^4} \qslash \right)
\,.
\ee
Note that in addition to the usual two constants, namely the wave 
function renormalization $Z$ and the bare mass $m$, we find 
two further integration constants $\nu_1^3$ and $\nu_2^4$ 
since the Gribov equation is a 
second order differential equation. According to their mass dimension 
one expects that they can be related (up to factors) to the quark and 
gluon condensates. The additional terms involving these two constants 
are singular at low $q^2$.  In QED one can simply drop these terms 
since the behavior of the Green function is perturbative at 
low momentum scales and the perturbative degrees of freedom 
in the Lagrangian coincide with the observable fields. 
In QCD the situation is drastically 
different since there are no freely propagating quarks at low 
momenta. Gribov's differential equation (\ref{gribgl}) hence relates 
the asymptotic solution (\ref{asymptfourterm}) at high momenta 
to a complicated 
and qualitatively different structure at low momenta. In order to obtain 
the full and correct solution at low momenta one must therefore in general 
keep all four terms of (\ref{asymptfourterm}) 
even in the regime of asymptotic freedom. 
The potential consequences of this result for applications of perturbative 
QCD still need to be explored. 

We now turn to the behavior of the dynamical mass function of the 
quark in the whole region of spacelike momenta in order to see 
how chiral symmetry breaking occurs \cite{Gribov:1998kb}. 
One finds from Gribov's equation (\ref{gribgl}) that the 
dynamical mass function can have oscillations and change its sign 
in an interval of momenta if the coupling constant exceeds its 
critical value (\ref{critcoupling}) in that 
interval \cite{Gribov:1991rv,Gribov:1998kb}. 
That critical value can be obtained analytically from (\ref{gribgl}). 
The full mass function can be calculated numerically for spacelike 
momenta \cite{Ewerz:2000qb}. 
Choosing the model (\ref{Kopplungma}) with $a=6$ for the 
running coupling one obtains\footnote{This figure has in fact been 
obtained using the modified equation 
(\protect\ref{gribglpion}) that includes pion corrections instead of the 
original equation  (\protect\ref{gribgl}). For spacelike momenta 
the mass functions resulting from the two equations have qualitatively 
the same behavior and their difference is rather small, 
see section \protect\ref{sec:pioneffects} below.}
the mass function $M(q^2)$ shown in 
Figure \ref{fig3m} for three different initial values at $q^2=0$. 
\begin{figure}[ht]
\centerline{\epsfxsize=8.6cm \epsfbox{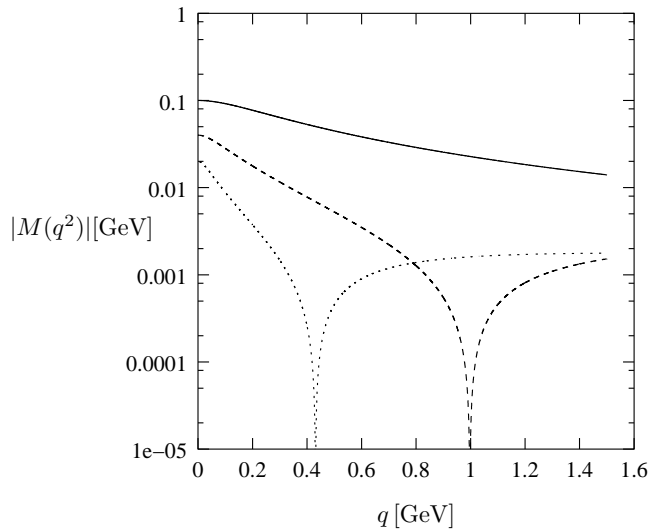}}   
\caption{The mass function for three different values of the renormalized mass 
\label{fig3m}}
\end{figure}
In the figure $q$ denotes the spacelike momentum, $q=\sqrt{-q^2}$. 
Depending on the initial value the mass function can stay 
positive (upper curve) or can have oscillations resulting in the spikes 
in the logarithmic plot in the two lower lines in Figure \ref{fig3m}. 

We now define the renormalized mass $m_R$ as the value of the 
dynamical mass function at vanishing quark momentum, 
\be
\label{defmR}
m_R = \lim_{q \to 0} M(q^2)
\,,
\ee
and further define a perturbative mass $m_P$ as the value of 
the mass function at a large (perturbative) momentum scale $\lambda$,  
\be
\label{defpertmass}
m_P= M(\lambda^2) 
\,. 
\ee
The precise value chosen for the scale $\lambda$ is not relevant for the 
present discussion as long as the coupling is no longer supercritical at 
that scale. In practice $\lambda$ can therefore be chosen as small 
as $2\,\mbox{GeV}$, for example. 

\begin{figure}[ht]
\centerline{\epsfxsize=8cm \epsfbox{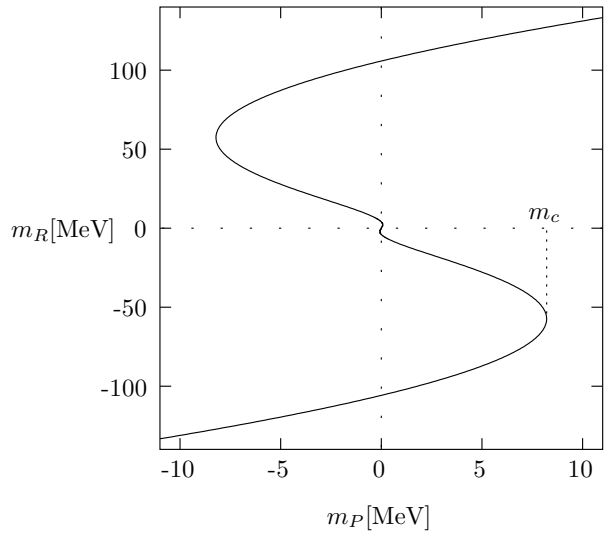}}   
\caption{The dependence of the renormalized mass $m_R$ on 
the perturbative mass $m_P$ reflecting chiral symmetry breaking 
\label{figm2m}}
\end{figure}

Recall that by solving the Gribov equation (\ref{gribgl}) one obtains a 
certain value for $m_P$ for each particular choice of $m_R$. 
We can hence consider the dependence of the renormalized mass $m_R$ 
on the perturbative mass $m_P$ (or vice versa). If the coupling 
constant is subcritical in the whole range of momenta (for example 
if $a$ in (\ref{Kopplungma}) is chosen sufficiently large) this dependence 
is one-to-one. If on the other hand the coupling is supercritical, the mass 
function can oscillate and different $m_R$ can correspond to the same 
$m_P$ at the scale $\lambda$. In this situation the dependence of $m_R$ 
on $m_P$ is no longer one-to-one and instead becomes multivalued, 
as is shown in Figure \ref{figm2m}. 
It can be easily seen that even for vanishing perturbative quark mass 
$m_P$ a large renormalized mass $m_R$ is generated. 
Hence chiral symmetry breaking occurs dynamically if the coupling 
is supercritical in the infrared. 

One can then proceed and compute the solutions to the Gribov equation 
(\ref{gribgl}) in the whole complex $q^2$-plane in order to study their 
analytic structure. In general this requires further assumptions about 
the behavior of $\alpha_s$ and hence of the gluon propagator $D_{\mu \nu}$, 
see (\ref{effphot}), for complex $q^2$. If one assumes that $D_{\mu \nu}$ 
does not have dramatic singularities off the real $q^2$-axis one finds that 
even for supercritical coupling the analytic structure of the quark's 
Green function does not correspond to a confined quark \cite{Ewerz:2004dd}. 
Instead, the Green function exhibits a pole and a cut on the positive real 
$q^2$-axis corresponding to the free propagation of the quark. 

\section{Pion Effects on the Quark's Green Function}
\label{sec:pioneffects}

In the process of chiral symmetry breaking pions are generated as 
Goldstone bosons and appear in the physical spectrum as massless 
states. Their Bethe-Salpeter amplitude $\phi_\pi$ is described by an 
equation \cite{Gribov:1998kb} that is derived 
in the same approximation as the equation (\ref{gribgl}) 
for the quark's Green function described above. For massless 
pions the solution of that equation can be found analytically in terms 
of the quark's Green function, 
\be
\label{bspion}
\phi_\pi \sim \{ \gamma_5, G^{-1} \}
\,.
\ee

Once pions appear as massless states in the physical spectrum due to 
chiral symmetry breaking they might have a considerable effect on the 
quark's Green function. This backreaction is not properly taken into 
account in the original equation (\ref{gribgl}). In order to obtain an 
improved equation for the Green function of light quarks one should 
therefore add their contribution separately \cite{Gribov:1999ui}. 
Incidentally, the pion propagator is proportional to $1/k^2$ such that 
one can treat pion loops on the quark using the same mathematical 
identity (\ref{deltafunktion}) as for the gluon propagator in Feynman gauge, 
hence again extracting the most IR-singular contributions. 
The coupling of the pion to the quark can be related to the pion decay 
constant $f_\pi$ via a Goldberger-Treiman relation, and taking into 
account the proper isospin factor for two light quark flavors one obtains 
the improved Gribov equation 
\be
\label{gribglpion}
\del^2 G^{-1} = g \,(\del^\mu G^{-1}) \,G \,(\del_\mu G^{-1}) 
- \frac{3}{16 \pi^2 f_\pi^2} 
\{i \gamma_5,G^{-1} \} \,G \, \{i \gamma_5,G^{-1} \} 
\,.
\ee
The pion decay constant $f_\pi$ can further be shown to fulfill in a 
selfconsistent way the equation 
\bea
\label{fpiselfcons}
f_\pi^2 &=& \frac{1}{8} \int \frac{d^4q}{(2 \pi)^4 i} \,
\mathrm{Tr} \left[ \{i \gamma_5,G^{-1} \} \,G \,\{i \gamma_5,G^{-1} \} 
\,G \, (\del_\mu G^{-1} G)^2 \right]
\nonumber \\
&&{}
+ \frac{1}{64 \pi^2 f_\pi^2} \int \frac{d^4q}{(2 \pi)^4 i} \,
\mathrm{Tr} \left[ \left( \{i \gamma_5,G^{-1} \} \,G \right)^4 \right] 
\,.
\eea
Note that also the modified Gribov equation (\ref{gribglpion}) 
is a differential equation rather than an integral equation and that it again 
involves only the quark's Green function. 

\begin{figure}[ht]
\centerline{\epsfxsize=8.6cm \epsfbox{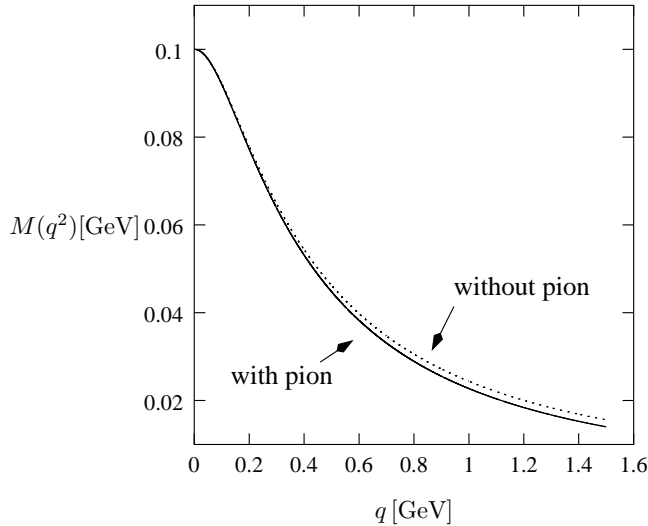}}   
\caption{Change in the mass function due to pion corrections 
\label{figmass}}
\end{figure}

Choosing for $f_\pi$ in (\ref{gribglpion}) its phenomenological value 
one can repeat the numerical analysis of the quark's Green function 
based on the improved Gribov equation \cite{Ewerz:tocome,Ewerz:2004dd}. 
One finds that for spacelike momenta the mass function $M(q^2)$ is 
affected only very little by the pion correction, as is illustrated in Figure 
\ref{figmass} for the choice $m_R=0.1 \,\mbox{GeV}$. 
As a consequence the general behavior of the Green function for 
spacelike momenta as it was found already with the original 
equation (\ref{gribgl}) remains the same when pion corrections 
are taken into account. In particular, the oscillations of the 
mass functions for supercritical coupling again lead to chiral 
symmetry breaking in exactly the same way as discussed above. 

Preliminary numerical results indicate that the situation is quite 
different when one moves away from spacelike momenta into 
the complex $q^2$-plane. Here the pion corrections become 
important and appear to induce considerable changes in the 
analytic structure of the Green function of light quarks. 
Gribov argued that the modified equation (\ref{gribglpion}) will lead 
to an analytic structure of the Green function corresponding to 
confined quarks and hence to the general confinement of color charges 
in QCD \cite{Gribov:1999ui}. 
A full numerical study of this conjecture remains to be carried out. 

\section{Summary and Outlook}
\label{sec:summary}

Gribov has developed an attractive physical picture of 
QCD in which chiral symmetry breaking and confinement 
are caused by the existence of light quarks which can trigger 
the mechanism of supercritical color charges. He has derived an 
equation for the Green function of light quarks that collects 
the most important contributions both in the infrared and in 
the ultraviolet momentum region. 
The equation describes chiral symmetry breaking 
and gives rise to pions as Goldstone bosons. There are some indications
that the backreaction of the pions on the quark leads to confinement. 

The next step in a systematic study of Gribov's approach should 
be a full investigation of the equation (\ref{gribglpion}) which includes 
pion corrections, in particular 
a study of the resulting analytic structure of the quark's Green function. 
A more difficult but extremely important future step should be 
to develop equations that describe the gluon sector of the theory in 
a similar approximation. Another crucial goal is to find observables 
that are sensitive to the particular confinement mechanism due to 
supercritical charges and to devise phenomenological tests of 
Gribov's scenario. 

In the course of his study of confinement Gribov has raised a number 
of questions concerning the general treatment of gauge theories in 
which the elementary fields of Lagrangian are not the ones that 
can propagate freely. In Gribov's picture a satisfactory solution of the 
confinement problem can only be obtained if these questions are 
addressed at the same time. That makes it a very challenging 
task to fully understand his ideas and to develop his program further 
in the future, but it is worth the effort. 

\section*{Acknowledgments}
I am very grateful that I had the opportunity to learn so much about 
the subject of this contribution and about quantum field theory in 
general from Vladimir Gribov in countless discussions and lectures 
when he was the supervisor of my diploma thesis. 
It was a great experience to see his unconventional way 
of thinking and his passion for physics from a close distance and 
to benefit from his deep understanding and insight.  

I would like to thank Julia Nyiri, Yuri Dokshitzer and P{\'e}ter L{\'e}vai 
for organizing this workshop and for inviting me to contribute to it. 
This work was supported by a Feodor Lynen fellowship of
the Alexander von Humboldt Foundation.

\end{document}